\documentclass[prd,showpacs,showkeys,floatfix,nofootinbib,
               preprint,
               fleqn,tightenlines]{revtex4}

\usepackage{amsmath}
\usepackage{amsfonts}
\usepackage{latexsym}
\usepackage{graphicx}
\usepackage{mathrsfs}
\usepackage{dcolumn}

\newcommand{\version}{v4}                      

\newcommand{\mathindentnum}{1cm}               

\newcommand{\fracnew}[2]
           {\protect\frac{{#1}_{\protect\vphantom{!_a}}}{{#2}^{\protect\vphantom{a}}}}

\newcommand{\widebar}{\overline}               
\newcommand{\beq}{\begin{equation}}
\newcommand{\eeq}{\end{equation}}
\newcommand{\beqa}{\begin{eqnarray}}
\newcommand{\eeqa}{\end{eqnarray}}

\newcommand{\threemat}[1]{\left( \begin{array}{ccc} #1 \end{array} \right)}

\begin{document}
\preprint{hep-ph/0509111 (\version)}
\vspace*{2\baselineskip}
\title{More on a possible energy dependence of $\Theta_{13}$ in
       vacuum neutrino oscillations}

\author{Frans R.\ Klinkhamer}
\email{frans.klinkhamer@physik.uni-karlsruhe.de}
\affiliation{Institute for Theoretical Physics,
University of Karlsruhe (TH), 76128 Karlsruhe, Germany\\ \\}

\begin{abstract}
\vspace*{.5\baselineskip}\noindent
Vacuum neutrino-oscillation probabilities from a
simple three-flavor model with both mass-square differences
and timelike Fermi-point splittings
have been presented in a previous article \mbox{[hep-ph/0504274].}
Here, further results are given: first, for specific parameters
relevant to MINOS in the low-energy mode and, then, for
arbitrary parameters. A generalized model with
equidistant Fermi-point splittings and an additional complex phase
is also discussed. If relevant, this generalized model might have
interesting effects at future long-baseline oscillation experiments.
\vspace*{2\baselineskip}
\end{abstract}


\maketitle

\section{Introduction}
\label{sec:Introduction}

In a previous article \cite{KlinkhamerTheta13}, we have considered
a simple three-flavor neutrino-oscillation model
with both mass-square differences ($\Delta m^2_{ij}$)
and timelike Fermi-point splittings ($\Delta b_0^{(ij)}$).
The mixing of the mass sector is taken to be bi-maximal and
the one of the Fermi-point-splitting sector trimaximal, with
all complex phases vanishing.
The model has furthermore a hierarchy of
Fermi-point splittings ($b_0^{(1)} = b_0^{(2)} \ne b_0^{(3)}$)
which parallels the hierarchy of masses ($m_1 = m_2 \ne m_3$).
This particular model may be called the ``stealth model,''
as it allows for Lorentz-noninvariant Fermi-point-splitting effects to
hide behind mass-difference effects. For the physics motivation of this type
of model discussion, see Ref.~\cite{KlinkhamerTheta13} and references therein.

The present article follows up the previous one in three ways.
First, we give model results of the vacuum appearance probability
$P(\nu_\mu\rightarrow \nu_e)$ relevant to MINOS
in the low-energy mode, complementing
the specific results for the medium energy mode
in Ref.~\cite{KlinkhamerTheta13}.

Second, we consider the model probability
$P_{\mu e} \equiv P(\nu_\mu\rightarrow \nu_e)$
for the case of relatively strong Fermi-point-splitting
effects compared to mass-difference effects,
whereas Ref.~\cite{KlinkhamerTheta13}
focused on relatively weak splitting effects. In particular,
we introduce a new parametrization (with nonnegative dimensionless parameters
$\rho$ and $\tau$) which makes a straightforward comparison
between different long-baseline neutrino-oscillation experiments possible.
Relatively weak or strong Fermi-point-splitting effects then correspond
to $\rho\tau \ll 1$ or $\rho\tau \gtrsim 1$, respectively.
The behavior of $P_{\mu e}(\rho,\tau)$ turns out to be quite
complicated for $\rho\tau \gtrsim 1$,

Third, we present results on the appearance probability
$P(\nu_\mu\rightarrow \nu_e)$ from a generalized model
with the same mass hierarchy as the model of
Ref.~\cite{KlinkhamerTheta13} but with equidistant Fermi-point splittings
($\,b_0^{(2)}- b_0^{(1)}=b_0^{(3)}- b_0^{(2)}\,$)
and one additional complex phase ($\omega =\pi/4$).
For completeness, we also give the model probability
of the time-reversed process, $\nu_e \rightarrow \nu_\mu$.
It will be seen that the generalized model has a rather interesting
phenomenology with stealth-like characteristics in certain cases
and strong time-reversal noninvariance in others.

The outline for the remainder of this article is as follows.
In Sec.~\ref{sec:Model}, we recall the definition of the model
and introduce the two neutrino-oscillation parameters $\rho$ and $\tau$.
In Sec.~\ref{sec:Results}, we present and discuss our new results.
In the Appendix, we consider one concrete model
and compare the capabilities of four accelerator-based
long-baseline oscillation experiments.

\section{Model}
\label{sec:Model}

Setting  $\hbar=c=1$ and writing $p \equiv |\vec p\,|$ for
the neutrino momentum, the Hamiltonian of the stealth model
\cite{KlinkhamerTheta13} contains
three terms in the $(\nu_e,\nu_\mu,\nu_\tau)$ flavor basis,
\begin{widetext}
\beq
H \supset p\,\openone +  X\cdot D_{m}\cdot X^{-1} + Y\cdot D_{b_0}\cdot Y^{-1} \,,
\label{3flavormatrix}
\eeq
with diagonal matrices
\beq
  D_{m}   \equiv   \mathrm{diag} \left(\,
\frac{m^2_{1}}{2 p} \, , \,
\frac{m^2_{2}}{2 p} \, , \,
\frac{m^2_{3}}{2 p}
\,\right) \, , \quad
 D_{b_0} \equiv  \mathrm{diag}\left(\,
b_0^{(1)}  \, , \,
b_0^{(2)}  \, , \,
b_0^{(3)}
\,\right) \, ,
\eeq
\mathindent=\mathindentnum
and $SU(3)$ matrices
\beqa
X &\equiv& M_{32}(\theta_{32}) \cdot M_{13}(\theta_{13},\delta)
           \cdot M_{21}(\theta_{21})
\,,\quad
Y \equiv M_{32}(\chi_{32}) \cdot M_{13}(\chi_{13},\omega)
         \cdot M_{21}(\chi_{21})\,,
\label{XYdef}
\eeqa
in terms of the basic  matrices
\beqa
M_{32}(\vartheta) &\equiv&
\threemat{1 &\;\;\; 0 \;\;\;& 0 \\
          0 &\;\;\; \cos\vartheta \;\;\;& \sin\vartheta \\
          0 &\;\;\; -\sin\vartheta \;\;\;&\cos\vartheta}
\,,\quad
M_{21}(\vartheta) \equiv
\threemat{\cos\vartheta &\;\;\; \sin\vartheta \;\;\;& 0 \\
          -\sin\vartheta&\;\;\; \cos\vartheta \;\;\;& 0 \\
         0 & 0 & 1}
\,,\nonumber\\[2mm]
M_{13}(\vartheta,\varphi) &\equiv&
\threemat{\cos\vartheta & \;\;\;0\;\;\; & \;+e^{+i\varphi}\,\sin\vartheta \\
          0 & \;\;\;1\;\;\; & 0 \\
          -e^{-i\varphi}\,\sin\vartheta& \;\;\;0\;\;\; & \cos\vartheta}
\,,
\label{Mdef}
\eeqa
and the following parameters:
\begin{subequations}
\label{fixedparameters}
\beqa
\Delta m^2_{21}    &\equiv& m_2^2-m_1^2 =0\,,\quad
R\equiv \Delta b_0^{(21)}/\Delta b_0^{(32)}
 \equiv (b_0^{(2)}- b_0^{(1)})/(b_0^{(3)}- b_0^{(2)})=0\,,
\label{fixedparameters-Delta21s} \\[4mm]
\theta_{13}&=&0\,,\quad\theta_{21}=\theta_{32}=
\chi_{13}=\chi_{21} =\chi_{32}=\pi/4\,,
\label{fixedparameters-mixingangles}\\[4mm]
\delta &=& \omega = 0\,.
\label{fixedparameters-phases}
\eeqa
\end{subequations}
With all other complex phases vanishing,
there are only two neutrino-oscillation parameters left in the model,
\beq
\Delta m^2_{31}   \equiv m_3^2-m_1^2           >0 \,,\quad
\Delta b_0^{(31)} \equiv b_0^{(3)}- b_0^{(1)}  >0\,,
\label{freeparameters}
\eeq
which have been taken positive.

For high-energy neutrino oscillations over a travel distance $L$, there
are, then, two dimensionless parameters
which completely define the problem, at least for the simple
model considered and matter effects  neglected.
These  neutrino-oscillation parameters can be defined as follows
($E_\nu \sim p$):
\begin{subequations} \label{rhotau}
\beqa
\rho &\equiv&
  \fracnew{2\,E_\nu\,\hbar c}{|\Delta m^2_{31}|\,c^4\,L} \approx
 \left( \fracnew{2.5\times 10^{-3}\;\mathrm{eV}^2/c^4}{|\Delta m^2_{31}|} \right)\,
 \left( \fracnew{735\;\mathrm{km}}{L} \right)\,
 \left( \fracnew{E_\nu}{4.656\;\mathrm{GeV}} \right)
 \,,
\label{rho}\\[3mm]
\newline
\tau  &\equiv&
  \fracnew{L\,|\Delta b_0^{(31)}|}{\hbar c} \approx
  \left( \fracnew{|\Delta b_0^{(31)}|}{2.685\times 10^{-13}\;\mathrm{eV}}\right)\,
  \left( \fracnew{L}{735\;\mathrm{km}} \right)   \,,
\label{tau}
\eeqa
\end{subequations}
with $\hbar$ and $c$ temporarily reinstated.

In terms of these parameters, an approximate formula for the
vacuum probability $P_{\mu e} \equiv  P(\nu_\mu\rightarrow \nu_e)$
is given by \cite{KlinkhamerTheta13}
\begin{subequations}\label{approxPandTheta13}
\beq
P_{\mu e}^\mathrm{\,stealth}(\rho,\tau) \sim
(1/2)  \,\sin^2 ( 2\,\Theta_{13} )\;
\sin^2 \left(\left[\: \rho^{-1} + \tau  +
\sqrt{\rho^{-2} + \tau^2} \;\right]/4\, \right) \,,
\label{approxP}
\eeq
with the following energy-dependent effective mixing angle:
\beq
\Theta_{13}(\rho,\tau) \sim (1/2) \,\arctan (\rho\,\tau)  \,.
\label{approxTheta13}
\eeq
\end{subequations}
Recall that, according to Eq.~\eqref{fixedparameters-mixingangles},
the bare mixing angle $\theta_{13}$ vanishes identically and that,
as mentioned above, matter effects are neglected in this approximate model
probability \cite{endnote-matter}.
\end{widetext}

More generally and allowing for a nonzero Fermi-point-splitting
ratio $R$, one can define
\begin{subequations}\label{defPandTheta13hat}
\beq
P_{\mu e}^\mathrm{\,stealth}(\rho,\tau) =
\sin^2 ( \widehat{\Theta}_{32} )  \,\sin^2 ( 2\,\widehat{\Theta}_{13} )\;
\widehat{\Lambda}_{\mu e}^{\,2} \,,
\label{defP}
\eeq
with the following functional dependence:
\beq
\widehat{\Theta}_{32}= \widehat{\Theta}_{32}(\rho\,\tau\,,R)  \,,\quad
\widehat{\Theta}_{13}= \widehat{\Theta}_{13}(\rho\,\tau\,,R)  \,,\quad
\widehat{\Lambda}_{\mu e}=\widehat{\Lambda}_{\mu e}(\rho\, ,\,\tau\,,R)\in [-1,+1]\,,
\label{defTheta13hat}
\eeq
\end{subequations}
so that the effective mixing angles $\widehat{\Theta}_{32}$ and
$\widehat{\Theta}_{13}$ are  independent of the travel
distance $L$, according to Eqs.~(\ref{rhotau}ab),
whereas $\widehat{\Lambda}_{\mu e}$ does depend on $L$ \cite{endnote-Lambdahat}.
In principle, $\widehat{\Theta}_{13}$
does not have to vanish in the limit $\rho \to 0$ for fixed $\tau$
and this is indeed the case for the generalized model
with splitting ratio $R \ne 0$ (see below).

\section{Results and discussion}
\label{sec:Results}

In Ref.~\cite{KlinkhamerTheta13}, we have given five figures
with model results, three of which may be relevant to T2K or NO$\nu$A and two
to MINOS in the medium-energy (ME) mode. Here, we present
one more figure, Fig.~\ref{fig01MINOS}, which may be relevant to
MINOS (baseline $L=735\;\mathrm{km}$) in the \emph{low}-energy (LE) mode,
with peak neutrino energy
$\widebar{E}_\nu$ approximately equal to $3.75\;\mathrm{GeV}$
and neutrino-oscillation length scale
$\widebar{L} \equiv 2\pi\widebar{E}_\nu/|\Delta m^2_{31}| \approx 1860 \;\mathrm{km}$
for $\Delta m^2_{31} = 2.5\times 10^{-3}\;\mathrm{eV}^2$.
Figure \ref{fig01MINOS}
(which may be compared to Figs. 4 and 5 of Ref.~\cite{KlinkhamerTheta13})
shows that, if MINOS--LE would be able to place an upper limit of $10\,\%$
on the appearance probability $P(\nu_\mu\rightarrow \nu_e)$
at the high end of the energy spectrum ($E_\nu \gtrsim 4\;\mathrm{GeV}$),
this would correspond to an upper limit on $|\Delta b_0^{(31)}|$
of approximately $3 \times 10^{-13}\;\mathrm{eV}$
(assuming $\tau \leq \pi$, see below).

A further figure, Fig.~\ref{fig02surface}, gives
numerical results which illustrate the general behavior of
the vacuum probability $P_{\mu e} \equiv  P(\nu_\mu\rightarrow \nu_e)$
as a function of the two dimensionless parameters $\rho$ and $\tau$
from Eqs.~(\ref{rhotau}ab). Note that we expect
$\lim_{\tau\to 0} P_{\mu e}(\rho,\tau)$ to vanish
[pure mass-difference model with $\Delta m^2_{21}=0$ and  $\theta_{13}=0$]
and $\lim_{\rho\to\infty} P_{\mu e}(\rho,\tau)$
to be given by $(1/2)\, \sin^2 (\tau/2)$ [pure Fermi-point-splitting model
with $\Delta b_0^{(21)}=0$, $\Delta b_0^{(31)}\ne 0$, and trimaximal mixing].
The landscape of Fig.~\ref{fig02surface} can then be described as follows:
mountain ridges start out at $\rho \gg 1$ and $\tau \approx n_\infty\,\pi$
for odd integers $n_\infty$, slope down towards lower values of $\rho$
keeping approximately the same values of $\tau$, and, finally,
disappear while bending
towards lower values of $\tau$ (more so for ridges with large $n_\infty$).
This topography is qualitatively reproduced by the analytic
expression (\ref{approxPandTheta13}).

Figure \ref{fig03tauslices} presents
a sequence of constant--$\tau$ slices of
the vacuum probability $P_{\mu e}(\rho,\tau)$ from Fig.~\ref{fig02surface}.
The behavior of $P_{\mu e}(\rho,\tau)$ at $\tau = 2\pi$
(or integer multiples thereof)
is quite remarkable, being nonzero only for a relatively small range
of energies; cf. the short-dashed curve in
the upper right panel of Fig.~\ref{fig03tauslices}.

Next, turn to a generalized stealth model with splitting
ratio $R=1$ (i.e., equidistant Fermi-point splittings)
and complex phase $\omega=0$ or $\pi/4$, the other model parameters being
kept at the values (\ref{fixedparameters}abc) and maintaining the
signs \eqref{freeparameters}.
(For high energies, this model is similar to the pure Fermi-point-splitting
model studied previously \cite{Klinkhamer-JETPL,Klinkhamer-IJMPA}.)
Figures \ref{fig04tauslices} and \ref{fig05tauslices} give the
resulting vacuum probabilities $P_{\mu e} \equiv P(\nu_\mu\rightarrow \nu_e)$
\cite{endnote-FPSmodels,endnote-deg-pert-theory}. For both complex phases,
the behavior of $P_{\mu e}(\rho,\tau)$ at $\tau \approx 12$
is particularly noteworthy; cf. the short-dashed curves in
the lower right panels of Figs.~\ref{fig04tauslices} and \ref{fig05tauslices}.
At the corresponding distance $L$
for given value of the Fermi-point splitting $\Delta b_0^{(31)}$, namely,
the stealth model lives up to its name by evading detection
via $\nu_e$ appearance, unless the experiment is able to reach down to
low enough neutrino energies ($\rho \sim 0.15$).
In principle, the way to corner this
stealth model would be to use several broad-band
experiments at \emph{different} baselines,
but this may require a substantial effort
(see the Appendix for a case study).

In the previous paragraph and corresponding Appendix,
the stealth-like behavior of the
$R=1$, $\omega=\pi/4$ model has been emphasized, but this holds
only for the $\nu_\mu\rightarrow \nu_e$ channel
relevant to superbeam experiments with an initial $\nu_\mu$  beam
from pion and kaon decays. In other channels, the
situation may be different. The probability $P_{e\mu}$ of the
$\nu_e \rightarrow \nu_\mu$ channel,
for example, is given in Fig.~\ref{fig06tauslices}.
The very different probabilities of Figs.~\ref{fig05tauslices}
and \ref{fig06tauslices}, for $\rho \gtrsim 0.2$ and generic values of $\tau$,
signal time-reversal (T) noninvariance.

For comparison, we give in Figs.~\ref{fig07tauslices}--\ref{fig010tauslices}
the numerical results from a ``realistic'' mass sector:
$R_m \equiv \Delta m^2_{21}/\Delta m^2_{32} \approx 0.033$,
$\sin^2 (2\theta_{13})= 0.1$
[close to the experimental bound from Chooz],
and two possible values of the complex phase, $\delta=0$ or $\delta=\pi/2\,$.
Figures \ref{fig09tauslices} and \ref{fig010tauslices}, in
particular, show that the maximum T--violation
discriminant $\Delta_T \equiv P_{e\mu}-P_{\mu e}$
from pure mass-difference neutrino oscillations ($\tau=0$) is
of the order of several percent, much less than the potential Fermi-point-splitting
result (several tens of percent) at the high-energy end of the neutrino spectrum.

The strong high-energy T violation
(and possibly CP violation \cite{Klinkhamer-IJMPA})
from Figs.~\ref{fig05tauslices}--\ref{fig010tauslices} traces
back to the large complex phase $\omega$ of the model considered, together
with the large mixing angles $\chi_{ij}$ and splitting ratio $R$ in the
Fermi-point-splitting sector. In other words, the breaking
of time-reversal invariance would primarily take place
outside the mass sector. A neutrino factory
(see Ref.~\cite{Bueno-etal2001} and the Appendix) would be
the ideal machine, in principle, to establish such strong T violation
in high-energy neutrino oscillations.

\section*{ACKNOWLEDGMENTS}

It is a pleasure to thank Jacob Schneps for useful discussions and
Elisabeth Kant and Christian
Kaufhold for help with the references and the figures.

\begin{appendix}
\section{Case study}
\label{sec:appendix}

In this appendix, we give a concrete example of the
complicated phenomenology of stealth--type models. Take, then,
the generalized model \eqref{3flavormatrix}--\eqref{freeparameters}
with parameters $\omega =\pi/4$ and
$\Delta b_0^{(31)} = 2\,\Delta b_0^{(21)}=
3.5 \times 10^{-12}\;\mathrm{eV}$
and consider the \emph{combined} performance of four accelerator-based
neutrino-oscillation experiments \cite{T'Jampens-NuFact04,Harris-NuFact04}:
the current K2K experiment (baseline $L=250\;\mathrm{km}$ and peak energy
$\widebar{E}_\nu \approx 1.0\;\mathrm{GeV}$),
the running MINOS experiment ($L=735\;\mathrm{km}$ and
$\widebar{E}_\nu \approx 3.75\;\mathrm{GeV}$, in the LE mode),
the planned T2K experiment ($L=295\;\mathrm{km}$ and
$\widebar{E}_\nu \approx 0.7\;\mathrm{GeV}$,
for  an off-axis beam at $2$ degrees),
and the  proposed NO$\nu$A experiment
($L=810\;\mathrm{km}$ and
$\widebar{E}_\nu \approx 2\;\mathrm{GeV}$,
at $14\;\mathrm{mrad}$ offset in the ME mode).
At the end of this appendix, we also comment briefly
on the capabilities of a possible neutrino factory.

First, calculate the model predictions for the two
current  experiments.
For K2K, the neutrino-oscillation parameters would be
$\tau \approx 4.4$ and $\rho \approx 0.63$, so that
a value for $P_{\mu e}\equiv P(\nu_\mu\rightarrow \nu_e)$
of only  $1\,\%$ would be expected
(cf. solid  curve in upper right panel of Fig.~\ref{fig05tauslices}),
which is consistent with the experimental result \cite{K2K-electron}.
[Note that the K2K appearance probability $P_{\mu e}$ would be approximately
$15\,\%$ for the model with vanishing complex phase $\omega$, as indicated by
the solid  curve in the upper right panel of Fig.~\ref{fig04tauslices}.]
For the MINOS--LE experiment with
$\tau \approx 13.0$ and $\rho \approx 0.81$,
a relatively small probability $P_{\mu e} \approx 7\,\%$ would be expected,
which would perhaps be hard to separate from the background.
[Note that this $\tau$--value corresponding to the MINOS baseline is an
order of magnitude above the largest one of Fig.~\ref{fig01MINOS}.]
Hence, the model predictions of the appearance probability
$P_{\mu e}$, for the chosen parameters,
would be consistent with rather low $\nu_e$ event rates from both of
these current experiments.
For completeness, the model would also give a survival probability
$P_{\mu \mu}\equiv P(\nu_\mu\rightarrow \nu_\mu)$
of some $49\,\%$ for K2K and $87\,\%$ for MINOS--LE.

Next, turn to the model predictions for the two future experiments
considered. For T2K with $\tau \approx 5.2$ and $\rho \approx 0.38$,
a substantial  $P_{\mu e}$ of some  $25\,\%$ would be expected
(cf. long-dashed curve in upper right panel of Fig.~\ref{fig05tauslices}).
Similarly, for NO$\nu$A with
$\tau \approx 14.4$ and $\rho \approx 0.39$,
$P_{\mu e} $ would be approximately $27\,\%$.
Hence, the model predictions of the appearance probability
$P_{\mu e}$ would imply
a clear signal in these future experiments,
at least for the chosen parameters.
Again, for completeness, the model survival probability $P_{\mu \mu}$ would be
approximately $71\,\%$ for T2K and  $20\,\%$ for NO$\nu$A.

Ultimately, the lowest values of $|\Delta b_0^{(31)}|$
could be probed by a neutrino factory with
broad energy spectrum $E_\nu \approx 10-50\;\mathrm{GeV}$ and
several detectors at baselines $L$ up to $12800\;\mathrm{km}$;
see, e.g., Refs.~\cite{Bueno-etal2001,CERNreportNuFact,NuFact04}.
For a quick estimate, one can simply compare
the expected experimental capabilities with
the approximate vacuum probability  (\ref{approxPandTheta13}),
for $\rho=\mathrm{O}(1)$ and $\tau \ll 1$ as defined by \eqref{rhotau}.
In the best of all worlds, a $P_{\mu e}$ sensitivity of $0.01\,\%$
at  $L=7350\;\mathrm{km}$ and $E_\nu \approx 46\;\mathrm{GeV}$,
for example, would correspond to $|\Delta b_0^{(31)}|$
of the order of $10^{-15}\;\mathrm{eV}$.
However, for a definite analysis at these large distances,
matter effects  \cite{endnote-matter} would need to be folded in.
\end{appendix}


\newpage
\setcounter{figure}{0}
\begin{figure*}[ht]
\begin{center}
\vspace*{-1cm}
\includegraphics[width=14cm]{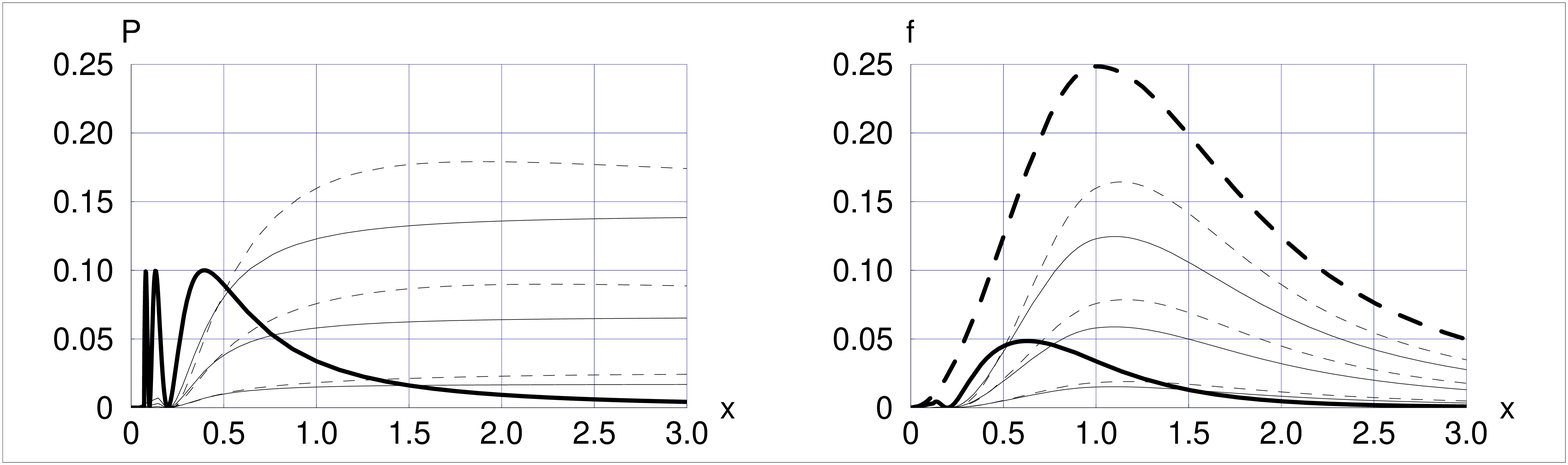}
\end{center}
\vspace*{-0.5cm}
\caption{\underline{Left:} Fixed-distance vacuum probabilities
$P_{\mu e} \equiv P(\nu_\mu\rightarrow \nu_e)$ at $l \equiv L/\widebar{L}
\equiv L\, |\Delta m^2_{31}| /(2\pi \widebar{E}_\nu)=735/1860$ vs.
dimensionless energy $x\equiv E_\nu/\widebar{E}_\nu$
for the ``stealth'' model of Ref.~\cite{KlinkhamerTheta13}
with both mass differences and timelike Fermi-point splittings, here
defined by Eqs.~\eqref{3flavormatrix}--\eqref{freeparameters}.
The  broken curves are given by the analytic
expression (4.4) of Ref.~\cite{KlinkhamerTheta13}
with parameters $\widebar{\Theta}_{13} \equiv
\widebar{E}_\nu\,|\Delta b_0^{(31)}|/|\Delta m^2_{31}|=0.15,\, 0.30, \,0.45$,
which correspond to Fermi-point splittings
$|\Delta b_0^{(13)}| \approx (1.0,\,2.0,\,3.0) \times 10^{-13}\;\mathrm{eV}$
for $\widebar{E}_\nu= 3.75\;\mathrm{GeV}$
and $\Delta m^2_{31}= 2.5\times 10^{-3}\;\mathrm{eV}^2$.
The same analytic expression is given here by
Eq.~ (\ref{approxPandTheta13}) with parameters
$\tau \approx (0.37,\, 0.74,\, 1.12)$.
These broken curves are only approximate
and the corresponding thin solid curves give the full numerical results.
The heavy solid curve gives, for comparison, the standard mass-difference
probability $P_{\mu e}^\text{MD}$
for $\Delta m^2_{21}=0$, $\sin^2 \theta_{32}=1/2$,
$\sin^2 (2\,\theta_{13})=0.2$, and $l=735/1860$.
\underline{Right:}
Electron--type neutrino energy spectra $f_e(x)$ at $l =735/1860$
from an initial muon--type spectrum (4.8) of Ref.~\cite{KlinkhamerTheta13}
at $l=0$ and vacuum probabilities $P(\nu_\mu\rightarrow \nu_e)$
of the left panel.
The initial $\nu_\mu$ spectrum multiplied by a constant factor $0.25$
is shown as the heavy long-dashed curve.
The resulting $\nu_e$ spectra from the ``stealth'' model
are shown as the thin solid curves [approximate values as thin broken curves].
The heavy solid curve gives,
for comparison, the $\nu_e$ spectrum from the standard mass-difference
probability $P_{\mu e}^\text{MD}$ of the left panel.}
\label{fig01MINOS}
\end{figure*}
\vspace*{2\baselineskip}
\begin{figure*}[hb]
\begin{center}
\vspace*{-1cm}
\includegraphics[width=14cm]{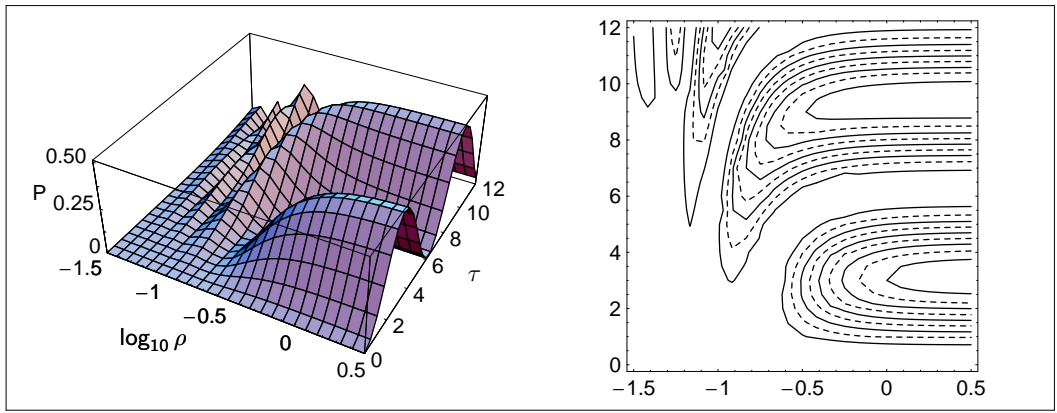}
\end{center}
\vspace*{-.5cm}
\caption{Numerical results for the
vacuum probability $P_{\mu e} \equiv P(\nu_\mu\rightarrow \nu_e)$
from the ``stealth'' model \eqref{3flavormatrix}--\eqref{freeparameters},
as a function of the two dimensionless parameters $\rho$ and $\tau$,
defined by Eqs.~(\ref{rhotau}ab).
\underline{Left:} surface plot.
\underline{Right:} contour plot, with equidistant
contours at $P_{\mu e}=0.05, 0.10, 0.15, \ldots\, 0.45$ and
contours at integer multiples of 0.10 shown dashed.}
\label{fig02surface}
\vspace*{-7cm}
\end{figure*}

\newpage
\begin{figure*}[ht]
\begin{center}
\vspace*{-1cm}
\includegraphics[width=14cm]{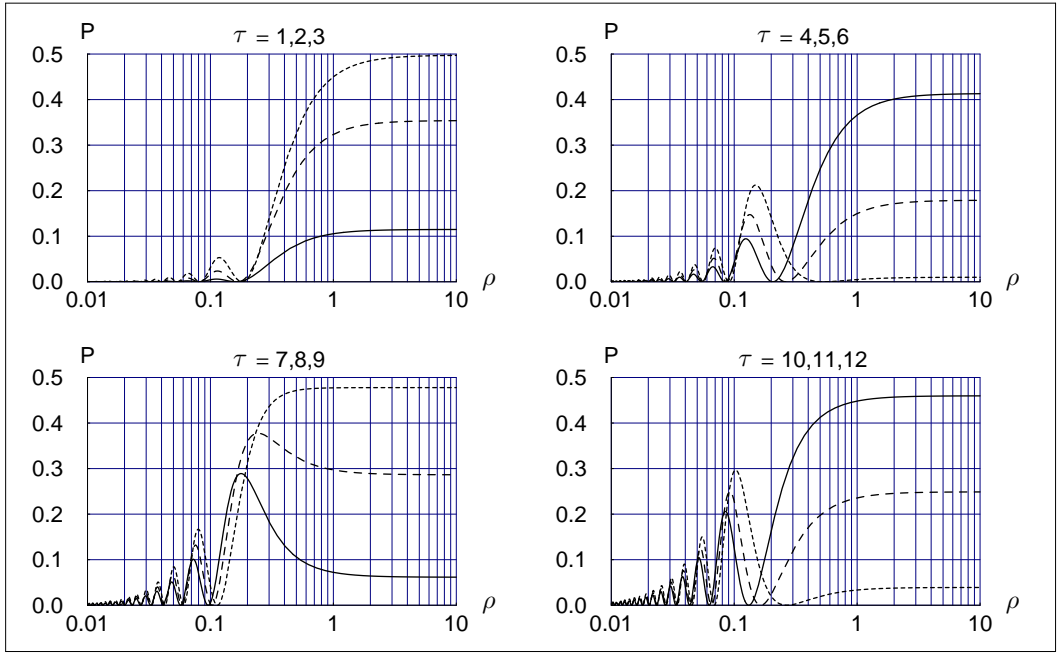}
\end{center}
\vspace*{-0.5cm}
\caption{Numerical results for constant--$\tau$ slices of the vacuum
probability $P \equiv P(\nu_\mu\rightarrow \nu_e)$ from Fig.~\ref{fig02surface}.
The curves for positive $\tau = 1,2,0 \pmod 3$
are shown as solid, long-dashed, and short-dashed lines, respectively.
For the simple model considered,
given by  Eqs.~\eqref{3flavormatrix}--\eqref{freeparameters},
the parameters $R$ and $\omega$ vanish identically.}
\label{fig03tauslices}
\vspace*{0cm}
\end{figure*}
\vspace*{2\baselineskip}
\begin{figure*}[hb]
\begin{center}
\vspace*{-1cm}
\includegraphics[width=14cm]{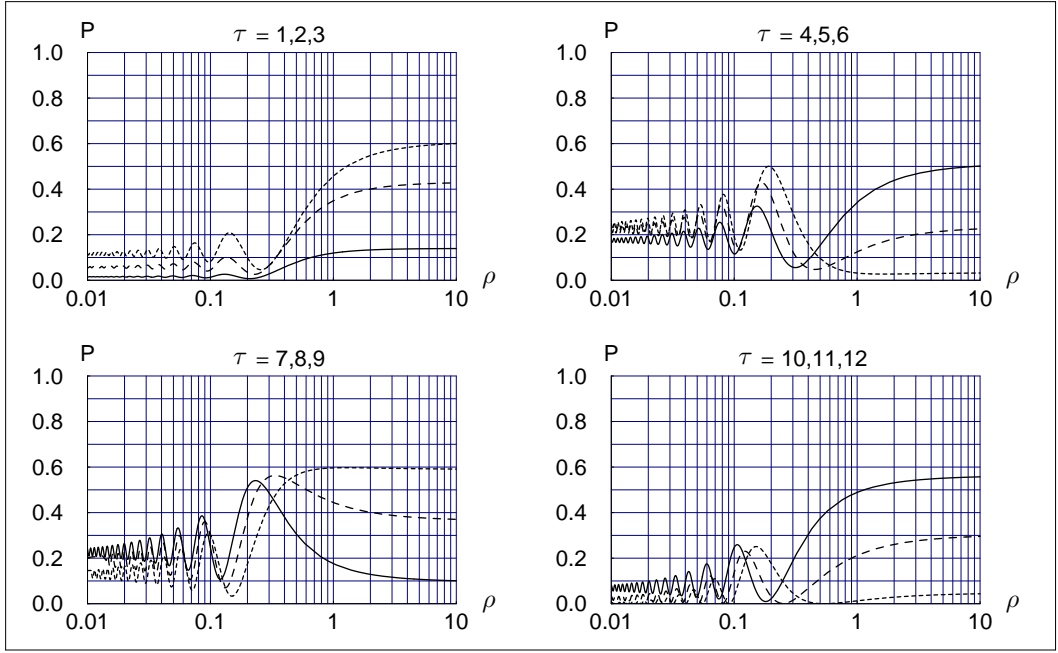}
\end{center}
\vspace*{-0.5cm}
\caption{Same as Fig.~\ref{fig03tauslices} but for
the generalized model with splitting ratio
$R\equiv |\Delta b_0^{(21)}/\Delta b_0^{(32)}| =1$,
the complex phase $\omega$ still vanishing.}
\label{fig04tauslices}
\vspace*{-7cm}
\end{figure*}

\newpage
\begin{figure*}[ht]
\begin{center}
\vspace*{-1cm}
\includegraphics[width=14cm]{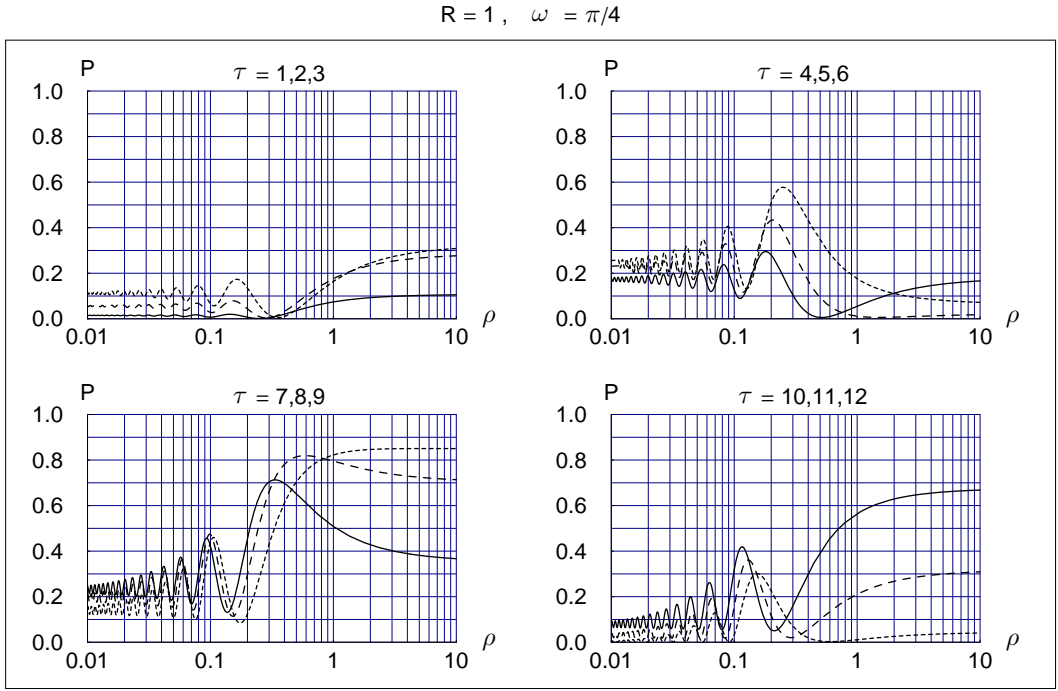}
\end{center}
\vspace*{-0.5cm}
\caption{Same as Fig.~\ref{fig04tauslices} but now for
the generalized model with complex phase $\omega =\pi/4$.}
\label{fig05tauslices}
\vspace*{-1.5cm}
\end{figure*}
\vspace*{7\baselineskip}
\begin{figure*}[hb]
\begin{center}
\vspace*{-1cm}
\includegraphics[width=14cm]{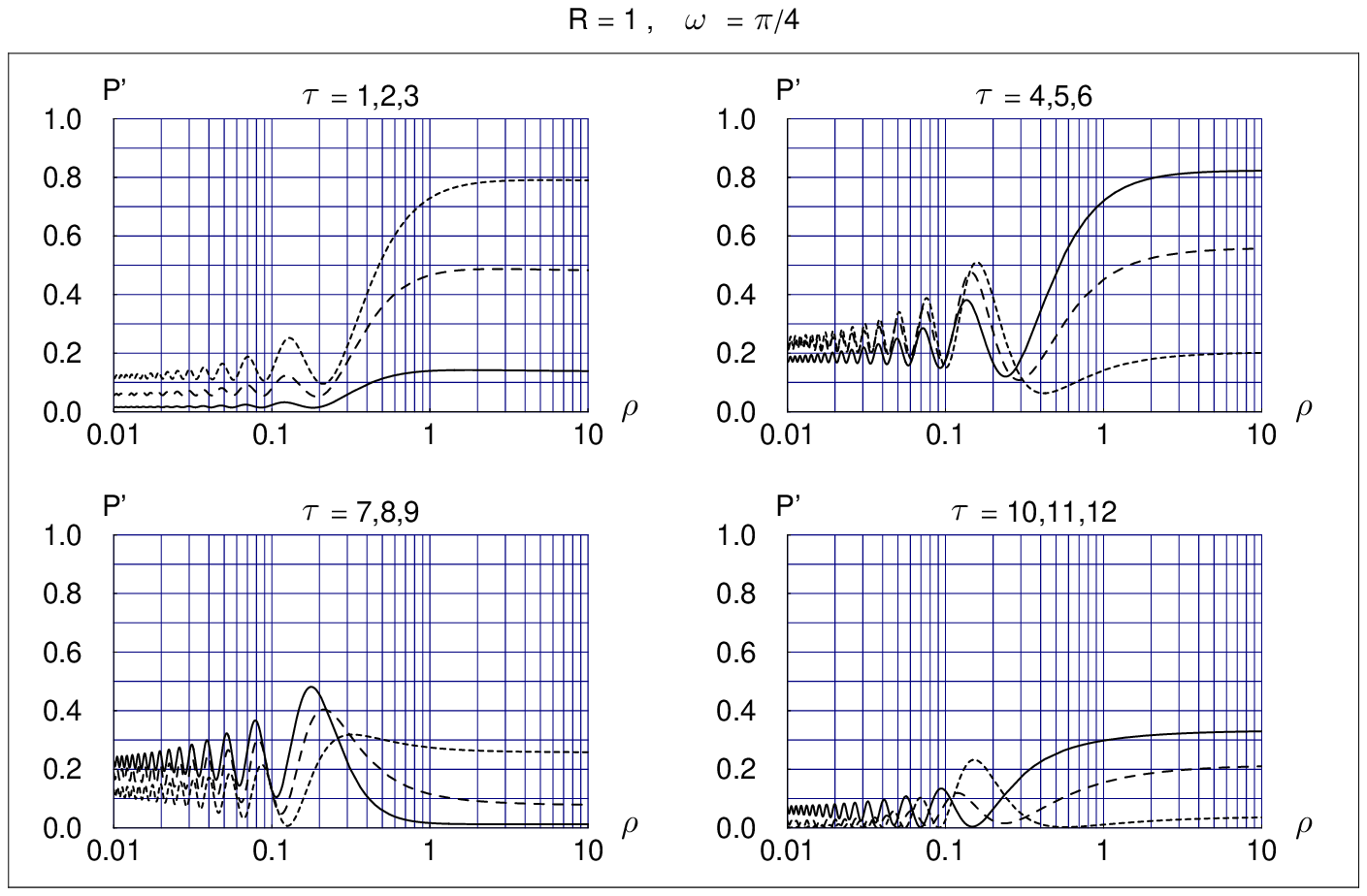}
\end{center}
\vspace*{-0.5cm}
\caption{Same as Fig.~\ref{fig05tauslices}
but for the time-reversed process, with probability
$P^\prime \equiv P(\nu_e\rightarrow \nu_\mu)$.
If CPT invariance holds, $P^\prime$ also corresponds
to $P(\overline{\nu}_\mu\rightarrow \overline{\nu}_e)$.}
\label{fig06tauslices}
\vspace*{-2cm}
\end{figure*}

\newpage
\begin{figure*}[ht]
\begin{center}
\vspace*{-1cm}
\includegraphics[width=14cm]{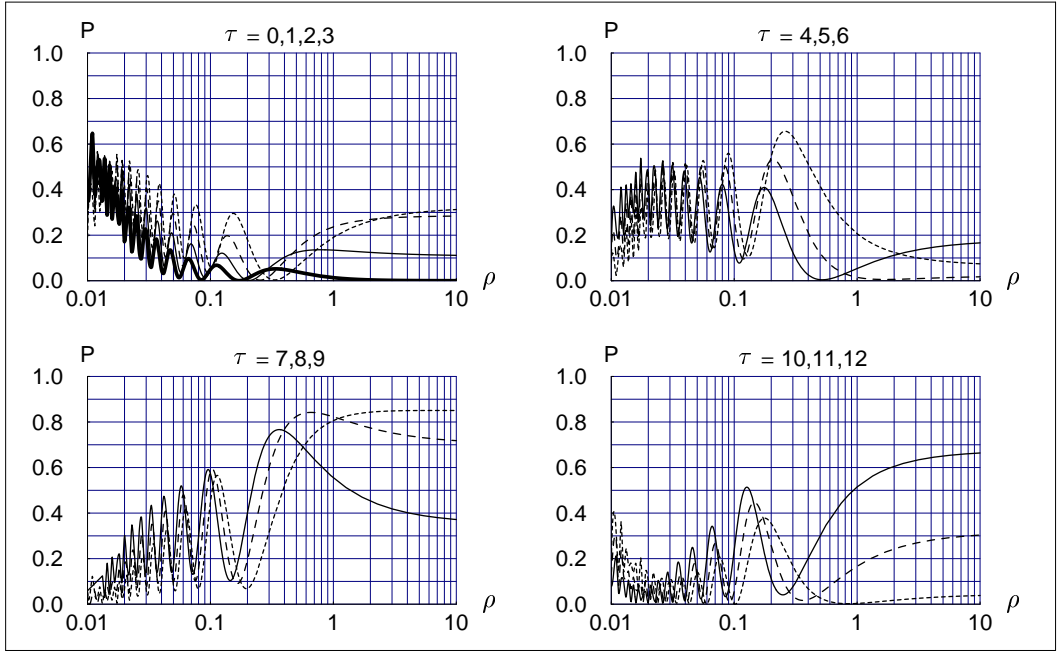}
\end{center}
\vspace*{-0.5cm}
\caption{Same as Fig.~\ref{fig05tauslices} but now for mass-sector
parameters $R_m \equiv \Delta m^2_{21}/\Delta m^2_{32} =1/30$,
$\sin^2 (2\theta_{13})= 1/10$, and $\delta=0$.
The heavy solid curve in the upper left panel has $\tau=0$,
corresponding to pure mass-difference neutrino oscillations.}
\label{fig07tauslices}
\end{figure*}
\vspace*{3\baselineskip}
\begin{figure*}[hb]
\begin{center}
\vspace*{-1cm}
\includegraphics[width=14cm]{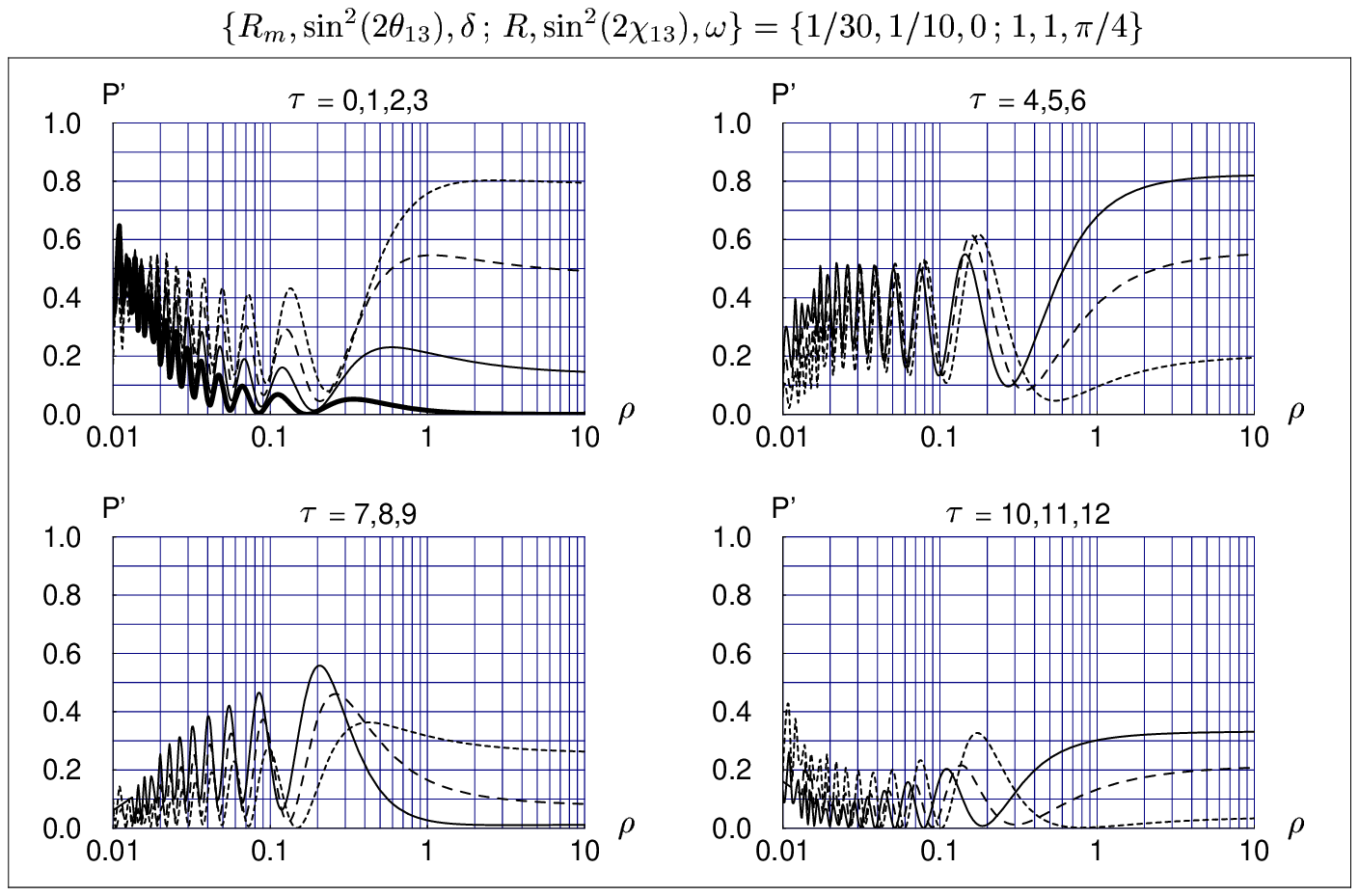}
\end{center}
\vspace*{-0.5cm}
\caption{Same as Fig.~\ref{fig07tauslices}  but for the time-reversed process,
with probability $P^\prime \equiv P(\nu_e\rightarrow \nu_\mu)$.
If CPT invariance holds, $P^\prime$ also corresponds
to $P(\overline{\nu}_\mu\rightarrow \overline{\nu}_e)$.}
\label{fig08tauslices}
\vspace*{-2cm}
\end{figure*}

\newpage
\begin{figure*}[ht]
\begin{center}
\vspace*{-1cm}
\includegraphics[width=14cm]{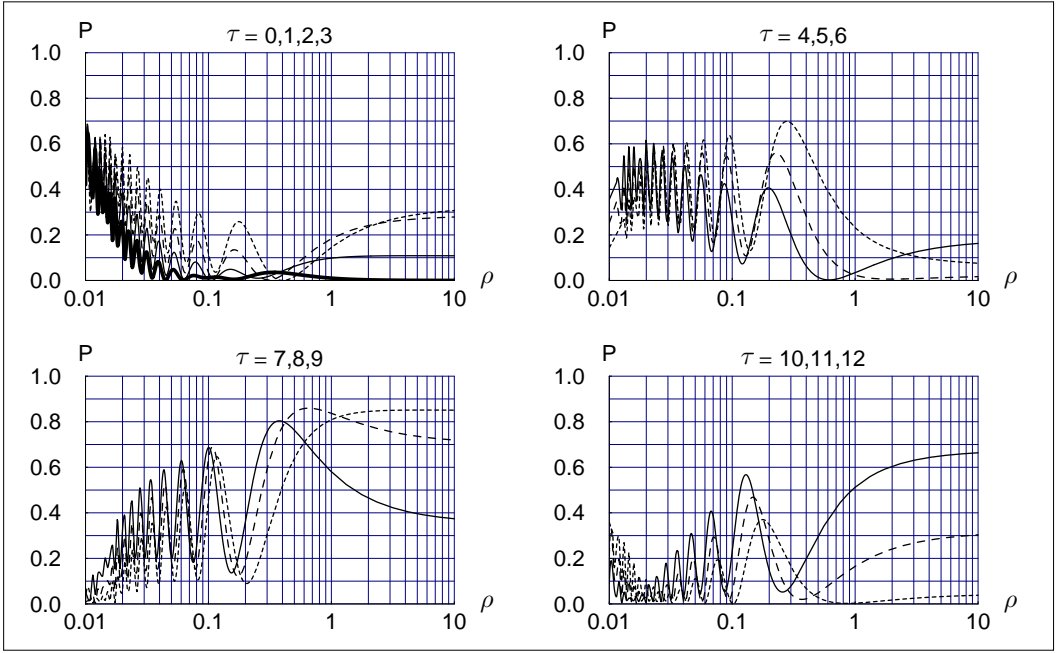}
\end{center}
\vspace*{-0.5cm}
\caption{Same as Fig.~\ref{fig07tauslices} but now for $\delta=\pi/2$.}
\label{fig09tauslices}
\end{figure*}
\vspace*{4\baselineskip}
\begin{figure*}[hb]
\begin{center}
\vspace*{-1cm}
\includegraphics[width=14cm]{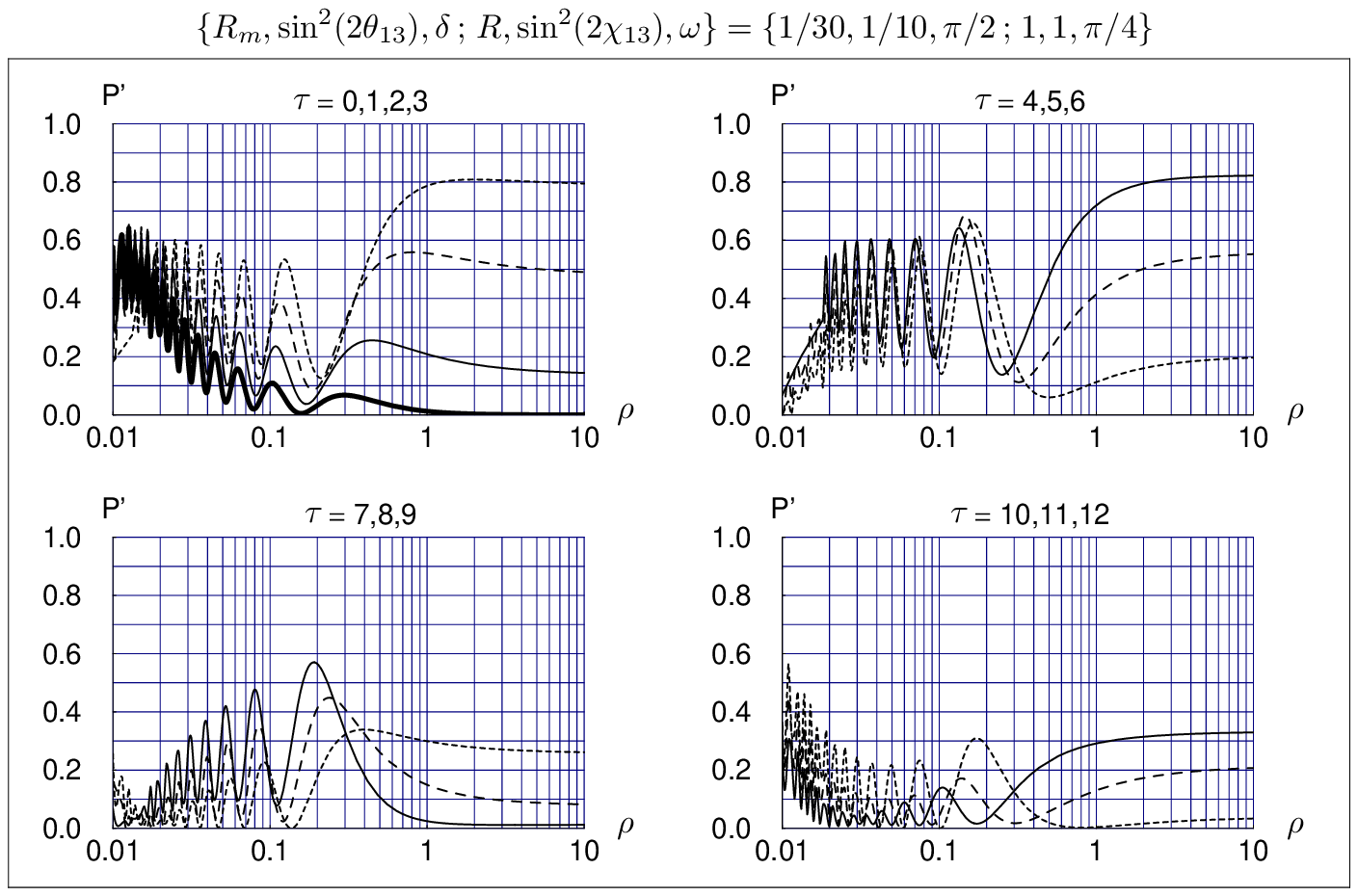}
\end{center}
\vspace*{-0.5cm}
\caption{Same as Fig.~\ref{fig09tauslices}  but for the time-reversed process,
with probability $P^\prime \equiv P(\nu_e\rightarrow \nu_\mu)$.
If CPT invariance holds, $P^\prime$ also corresponds
to $P(\overline{\nu}_\mu\rightarrow \overline{\nu}_e)$.}
\label{fig010tauslices}
\vspace*{-2cm}
\end{figure*}


\begin{thebibliography}{9}

\bibitem{KlinkhamerTheta13}
 F.R. Klinkhamer,
``Possible energy dependence of $\Theta_{13}$ in neutrino oscillations,''
Phys. Rev. D {\bf 71}, 113008 (2005) [hep-ph/0504274].

\bibitem{endnote-matter}
Matter effects (coherent forward scattering) from the Earth's mantle
become important for standard mass-difference neutrino oscillations
at energies $E_\nu$ of order $10\;\mathrm{GeV}$
and travel distances $L$ of order $2500\;\mathrm{km}$.
For further details and references, see, e.g.,
Ref.~\cite{Bueno-etal2001}
and Ref.~\cite{CERNreportNuFact}, Chapter 3 [hep-ph/0210192].


\bibitem{Bueno-etal2001}
A. Bueno, M. Campanelli, S. Navas-Concha, and A. Rubbia,
``On the energy and baseline optimization to study effects related to the
$\delta$--phase (CP--/T--violation) in neutrino oscillations at a neutrino factory,''
Nucl.\ Phys.\ B {\bf 631}, 239 (2002) [hep-ph/0112297].


\bibitem{CERNreportNuFact}
A. Blondel {\it et al.},
\emph{ECFA/CERN studies of a European neutrino factory complex},
report CERN-2004-002, April 2004.

\bibitem{endnote-Lambdahat}
In the combined limit $\rho\to \infty$ and $\tau\to 0$, the factor
$\widehat{\Lambda}_{\mu e}$ behaves as $\mathrm{O}(1/\rho)+\mathrm{O}(\tau)$,
i.e., proportional to $L\,$. Note also that, in first approximation,
$\widehat{\Lambda}_{\mu e}$ is assumed to contain no further
factors which only depend on the product $\rho\tau$.
In addition, it is necessary to consider all channels in order to make,
as far as possible, a useful distinction between the quantities
$\widehat{\Theta}_{32}$ and $\widehat{\Theta}_{13}$ in Eq.~\eqref{defP}.


\bibitem{Klinkhamer-JETPL}
F.R. Klinkhamer,
\emph{Neutrino oscillations from the splitting of Fermi points},
JETP Lett. {\bf 79}, 451 (2004) [hep-ph/0403285].

\bibitem{Klinkhamer-IJMPA}
F.R. Klinkhamer,
\emph{Lorentz-noninvariant neutrino oscillations: Model and predictions},
Int. J. Mod. Phys.  A  {\bf 21}, 161 (2006) [hep-ph/0407200].

\bibitem{endnote-FPSmodels}
Note that the notation for the neutrino dispersion relation
in Refs.~\cite{Klinkhamer-JETPL,Klinkhamer-IJMPA}
differs from the one used here and in
Ref.~\cite{KlinkhamerTheta13}. For neutrino oscillations
in the pure Fermi-point-splitting model \cite{Klinkhamer-JETPL,Klinkhamer-IJMPA},
the different sign of $b_0^{(f)}$ can be compensated by a
sign change of the complex phase $\epsilon$.
For example, setting $\epsilon=-\omega$ in the exact model probablity
$P_{\mu e}$ from Eq.~(11e) of Ref.~\cite{Klinkhamer-IJMPA}
reproduces the numerical $\rho \to \infty$ results for
$R=1$, $\chi_{21}=\chi_{32}=\pi/4$,
$\chi_{13}=\arctan\sqrt{1/2}\approx \pi/5$, and $\omega=0$ or $\pi/4$
(these numerical results closely resemble those for $\chi_{13}=\pi/4$
presented in Figs.~\ref{fig04tauslices} and \ref{fig05tauslices} here).


\bibitem{endnote-deg-pert-theory}
Degenerate perturbation theory for $\rho \to 0$
and fixed $\tau$ (arbitrary $R$ and $\omega$) gives
$P_{\mu e} \sim (4 R^2/S)\, \sin^2 \,[\sqrt{S}\,\tau/(8+8R)]$
with $S \equiv 4+4 R +9 R^2$.
For $R=1$, in particular, the $\rho \to 0$ probablity is given by
$P_{\mu e} \sim (4/17)\, \sin^2 \,[\sqrt{17}\,\tau/16 ]$.
If, however,  the mass-hierarchy parameter
$R_m \equiv \Delta m^2_{21}/\Delta m^2_{32}$
is changed from zero  to $1/30$, the $\rho \lesssim 0.1$  behavior
changes significantly, as will become clear later on.


\bibitem{T'Jampens-NuFact04}
S. T'Jampens,
\emph{Current and near future long-baseline neutrino experiments},
in Ref.~\cite{NuFact04},  p. 24.

\bibitem{Harris-NuFact04}
D.A. Harris, \emph{Superbeam Experiments},
in Ref.~\cite{NuFact04}, p.  34.

\bibitem{NuFact04}
\emph{Proceedings 6th International Workshop on Neutrino Factories
and Superbeams (NuFact04)}, edited by M. Aoki, Y. Iwashita, and M. Kuze,
Nucl. Phys. B (Proc. Suppl.) {\bf 149}  (2005), pp. 1--411.



\bibitem{K2K-electron}
M.H. Ahn {\it et al.}  [K2K Collaboration],
\emph{Search for electron neutrino appearance in a 250-km long-baseline
experiment},
Phys.\ Rev.\ Lett.  {\bf 93}, 051801 (2004), hep-ex/0402017;
K. Kaneyuki, \emph{K2K far detector analysis},
in Ref.~\cite{NuFact04}, p.  119.



\end{thebibliography}
\end{document}